\documentclass[11pt, a4paper]{article}
\usepackage{moriond,epsfig}

\usepackage{epsf,color,latexsym}

\usepackage[latin1]{inputenc}

\def\siml{{\ \lower-1.2pt\vbox{\hbox{\rlap{$<$}\lower6pt\vbox{\hbox{$\sim$}}}}\ }} 

\newcommand{\als}{\alpha_s}

\newcommand{\nn}{\nonumber}

\def\be{\begin{equation}}
\def\ee{\end{equation}}
\def\bea{\begin{eqnarray}}
\def\eea{\end{eqnarray}}

\begin{document}
\vspace*{4cm}
\title{Non-relativistic Quantum Mechanics versus Quantum Field Theories}

\author{Antonio Pineda}
\address{Grup de Física Teòrica and IFAE\\
Universitat Autònoma de Barcelona, E-08193 Bellaterra (Barcelona), Spain}
%

\maketitle\abstracts{We briefly review the 
derivation of a non-relativistic quantum mechanics description of 
a weakly bound non-relativistic system from the underlying quantum field 
theory. We highlight the main techniques used.}

In a first approximation, the dynamics of the Hydrogen atom can be described 
by the solution of the Schrödinger equation with a Coulomb potential.
However, it is not always clear how to derive this equation 
from the more fundamental quantum field theory, QED, much less how 
to get corrections in a systematic way. A similar problem is faced in  
heavy quarkonium systems with very large heavy quark masses. In this situation 
the dynamics is mainly perturbative and 
one efficient solution to this problem comes from the use of 
effective field theories (EFTs) and in particular of pNRQCD\cite{Pineda:1997bj}\footnote{For a comprehensive 
review of pNRQCD see\cite{Brambilla:2004jw}.}. This EFT takes full advantage of the hierarchy of scales that appear 
in the system ($v$ is the velocity of the heavy quark in the center of mass frame 
and $m$ is the heavy quark mass):
\be
\label{hierarchy}
m \gg mv \gg mv^2 \cdots
\ee 
and makes systematic and natural the connection of the Quantum Field Theory 
with the Schrödinger equation. Roughly speaking the EFT turns out to be 
something like:
\begin{eqnarray*}
\,\left.
\begin{array}{ll}
&
\displaystyle{\left(i\partial_0-{{\bf p}^2 \over 2m}-V_s^{(0)}(r)\right)\Phi({\bf r})=0}
\displaystyle{{\rm \ + \ corrections\; to\; the\; potential}}
\\
&
\displaystyle{+{\rm interaction \;with\; other\; low}-{\rm energy\; degrees \;of\; freedom}}
\end{array} \right\} 
{\rm pNRQCD}
\end{eqnarray*}
where $V_s^{(0)}(r)=-C_f\als/r$ in the perturbative case and $\Phi({\bf r})$ is the
${\bar Q}$--$Q$ wave-function. This EFT\footnote{It is also possible to 
study heavy quarkonium 
systems in the non-perturbative regime with pNRQCD 
profiting from the hierarchy of scales of Eq. (\ref{hierarchy}), see\cite{Brambilla:2000gk,Pineda:2000sz}.} 
is relevant, at least, for the study of the 
ground state properties of the bottomonium system, non-relativistic sum rules 
and the production of $t$-$\bar t$ near threshold (for some recent applications 
see\cite{Beneke:2005hg,Penin:2005eu,Pineda:2006gx,Pineda:2006ri,Brambilla:2007cz}).

The key point in the construction of the EFT is to determine 
the kinematic situation we want to describe.  This fixes the (energy of the) degrees of freedom that appear as 
physical states (and not only as loop fluctuations). In our case the degrees of freedom in pNRQCD are kept to have 
{\color{red}$E \sim mv^2$}. In order to derive pNRQCD we sequentially integrate out the larger scales.
\begin{figure}[h]
\hspace{-0.3in}
\put(174,83){\cite{Caswell:1985ui}}
\put(230,5){{\color{red}\Large $E \sim mv^2$}}
\epsfxsize=3.8in
\centerline{\epsffile{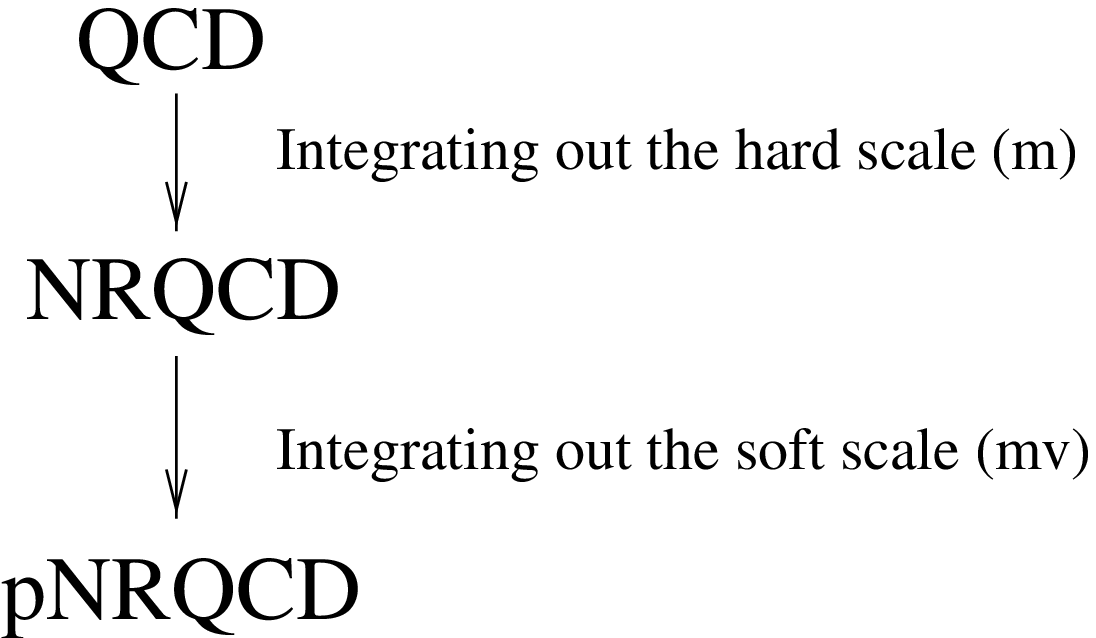}}
\end{figure}

In this paper, we would like to highlight  
the main techniques needed in order 
to perform efficiently high-precision perturbative computations 
in non-relativistic bound state systems. They can be summarized in four points:    
{\color{red}
\begin{enumerate}
\item
Matching QCD to NRQCD: Relativistic Feynman diagrams
\item
Matching NRQCD to pNRQCD (getting the potential): 
Non-Relativistic (HQET-like) Feynman diagrams
\item
Observable: Quantum mechanics perturbation theory 
\item
Observable: Ultrasoft loops 
\end{enumerate}}
The first two points explain the techniques needed to obtain pNRQCD 
from QCD, whereas the last two explain the kind of computations faced in 
the EFT when computing observables. All the computations can be performed in dimensional regularization 
and only one scale appears in each type of integral, which becomes homogeneous. 
This is a very strong simplification 
of the problem. In practice this is implemented in the following way:

{\color{red}Point 1)}.  One analytically expands over the three-momentum and residual energy
in the integrand before the integration is made in both the full and
the effective theory\cite{Manohar:1997qy,Pineda:1998kj}. 
\bea
\nn
{\color{red}QCD} \qquad
\int d^4q f(q,m,|{\bf p}|,E)&=& \int d^4q f(q,m,0,0)+
{\cal O}\left({E \over m},
{|{\bf p}|\over m} \right)
\sim C(\frac{\mu}{m})({\rm tree\ level})|_{NRQCD}
\\
{\color{red}NRQCD} \qquad
 \int d^4q f(q,|{\bf p}|,E)&=& \int d^4q f(q,0,0)={\color{red}0\,!!}
\eea
Therefore, the computation of loops in the effective theory just gives \textcolor{red}{zero} and 
{\color{red}
one matches loops in QCD with only one scale (the mass) to tree level diagrams in NRQCD}, which we 
schematically draw in the following figure:
\medskip
\begin{figure}[h]
\hspace{-.1in}
\epsfxsize=5in
\centerline{\epsffile{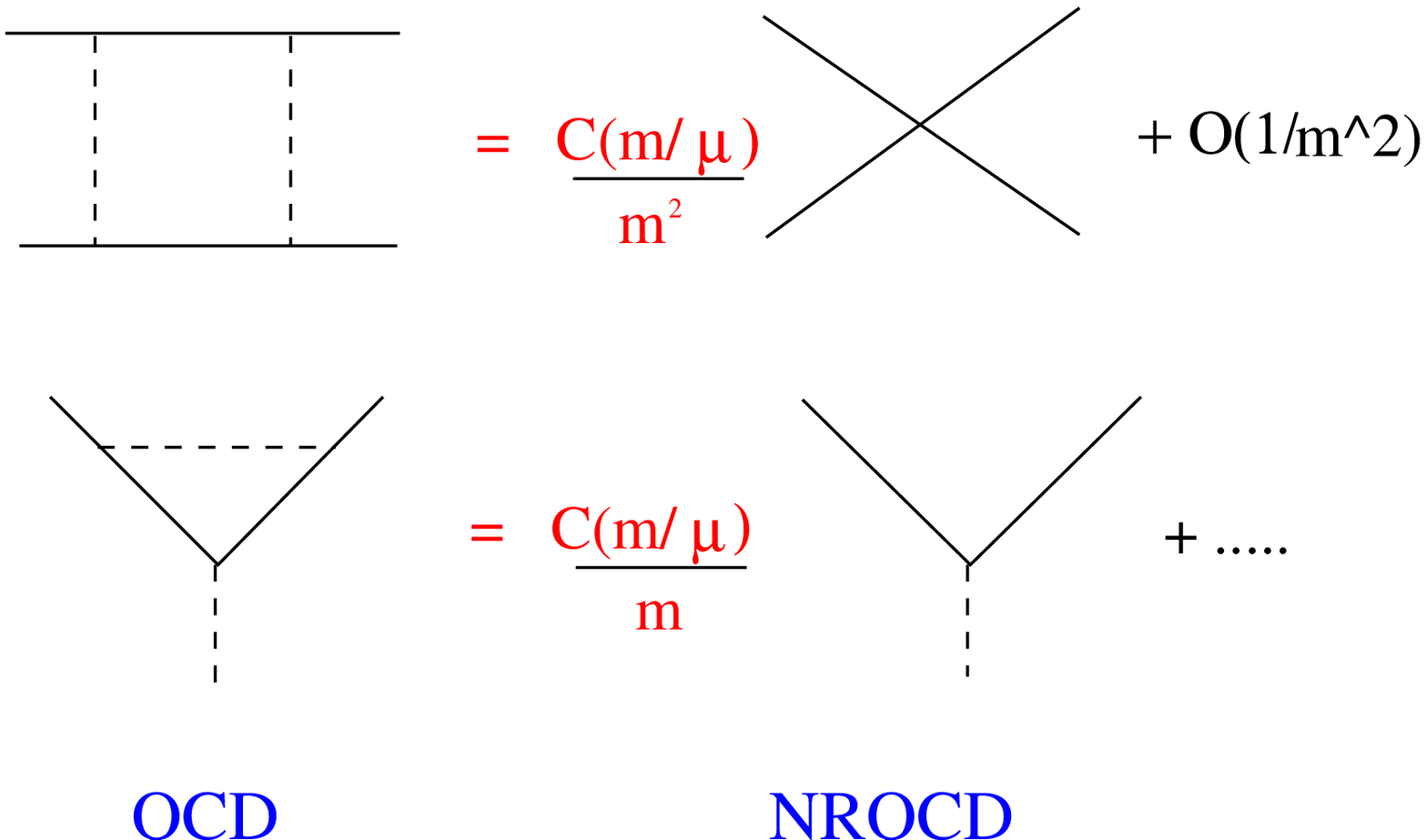}}
\end{figure}

\vfill
\newpage

{\color{red}Point 2)} works analogously\cite{Pineda:1998kn}. One expands in the
scales that are left in the effective theory. We integrate out the
scale {\color{red} k} (transfer momentum between the quark and antiquark). 
Again loops in the EFT are zero and only tree-level 
diagrams have to be computed in the EFT: 
\bea
{\color{red}NRQCD}  \qquad \int d^4q f(q,k,|{\bf p}|,E)&=& 
\int d^4q f(q,k,0,0)+{\cal O}\left({E \over k},
{|{\bf p}|\over k} \right) \sim \delta h_s ({\rm potential})
\\
{\color{red}pNRQCD}
\qquad
\int d^4q f(q,|{\bf p}|,E)&=& \int d^4q f(q,0,0)={\color{red}0\,!!}
\eea
We illustrate the matching in the figure below.  Formally the one-loop diagram 
is equal to the QCD diagram shown above. The difference is that it  
has to be computed with the HQET quark propagator ($1/(q^0+i\epsilon)$) and 
the vertices are also different.
\begin{figure}[htb]
\makebox[0.0cm]{\phantom b}
\put(100,153){${\bf p}$}\put(120,162){$>$}
\put(170,153){${\bf p}'$}\put(190,162){$>$}
\put(155,130){${\bf k}={\bf p}-{\bf p}'$}
\put(100,100){\epsfxsize=10truecm \epsfbox{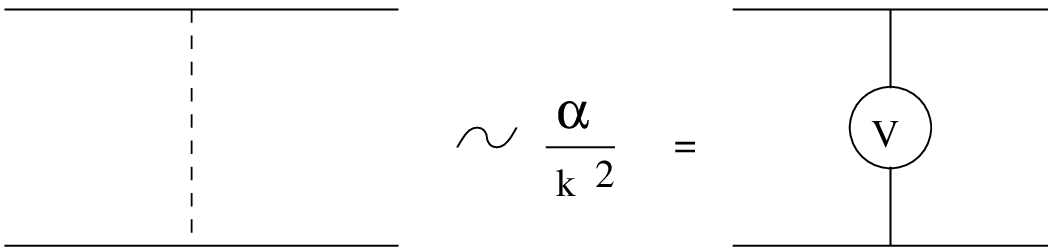}}
\put(100,20){\epsfxsize=10truecm \epsfbox{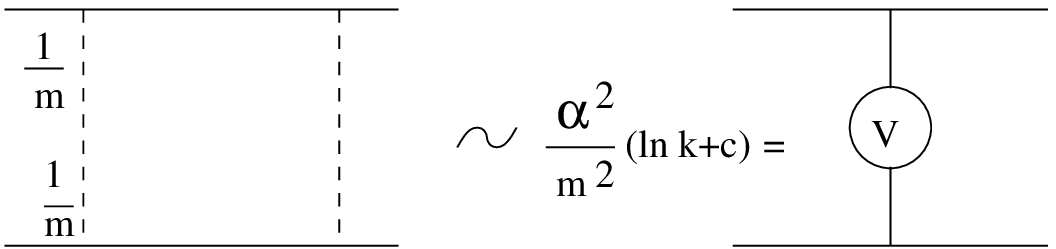}}
\put(145,0){NRQCD}
\put(320,0){pNRQCD}
\vspace{1mm}
\end{figure}

Once the Lagrangian of pNRQCD has been obtained one can 
compute observables. A key quantity in this respect is 
the Green function. In order to go beyond the leading order 
description of the bound state one has to compute 
corrections to the Green Function 
($\delta h_s$ schematically represents the corrections to the 
potential and $H_I$ the interaction with ultrasoft gluons):
$$
G_s(E)={1 \over h^{(0)}_s+\delta h_s-H_I-E}=G_s^{(0)}+\delta G_s
\qquad
G_s^{(0)}(E)={1 \over \displaystyle{h_s^{(0)}-E}}
\ .
$$
These corrections can be organized as an expansion
in {\color{red}$1/m$}, {\color{red}$\als$} and the {\color{red}multipole
  expansion}. Two type of integrals appear then, which correspond to points 3) and 4) above.

\vfill
\newpage
  
\textcolor{red}{\bf Point 3)}. For example, if we were interested in computing the spectrum at $O(m\als^6)$ 
(for QED see\cite{Czarnecki:1999mw}), one
should consider the iteration of subleading potentials ($\delta h_s$) in the
propagator:
\begin{figure}[htb]
\makebox[0.0cm]{\phantom b}
\put(10,1){$\displaystyle{\delta G_s^{pot.}=}$}
\put(60,1){\epsfxsize=3truecm \epsfbox{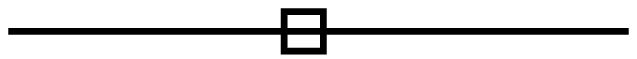}}
\put(180,1){\epsfxsize=3truecm \epsfbox{pot_insert.eps}}
\put(220,1){\epsfxsize=3truecm \epsfbox{pot_insert.eps}}
\put(93,12){$\delta h_s$}
\put(213,12){$\delta h_s$}
\put(253,12){$\delta h_s$}
\put(320,1){$+ \cdots$}
\put(160,1){$+$}
\put(30,-24){$\displaystyle{\sim {1 \over h_s^{(0)}-E}\delta h_s {1 \over h_s^{(0)}-E}
+
{1 \over h_s^{(0)}-E}\delta h_s {1 \over h_s^{(0)}-E}\delta h_s {1 \over h_s^{(0)}-E}
+\cdots}$}
\end{figure}

At some point, these corrections produce divergences. For example, a correction of the 
type: $\delta(r)G_s^{(0)}(C_f\als/r)G_s^{(0)}\delta(r)$, would produce the following divergence
\bea
&&
\langle{\bf r}=0|
{1 \over \displaystyle{E-{\bf p}^2/m}}
C_f {\als \over r}
{1 \over \displaystyle{E-{\bf p}^2/m}}
|{\bf r}=0\rangle
\nn
\\
\nn
&&
\qquad
\sim 
\int \frac{
{\rm d}^d p' }{ (2\pi)^d } \int \frac{ {\rm d}^d p }
{ (2\pi)^d } \frac{ m }{{\bf p}'^2 - mE } 
C_f
\frac{ 4\pi\als }{ ({\bf p-p'})^2 } \frac{ m }{{\bf p}^2-m E } 
\sim
- C_f\frac{m^2\als}{16\pi}  
\left(\frac{ 1 }{\epsilon }+2\ln(\frac{mE}{\mu_p})+\cdots\right).
\eea
Nevertheless, the existence of divergences in the effective theory is not a problem since 
they get absorbed in the potentials ($\delta h_s$). The same happens 
with ultrasoft gluons, {\color{red} \bf point 4)}\cite{Pineda:1997ie,Kniehl:1999ud}:
\begin{figure}[htb]
\makebox[0.0cm]{\phantom b}
\put(0,1){$\delta G_s^{\rm us}=$}
\put(50,-2){\epsfxsize=5truecm \epsfbox{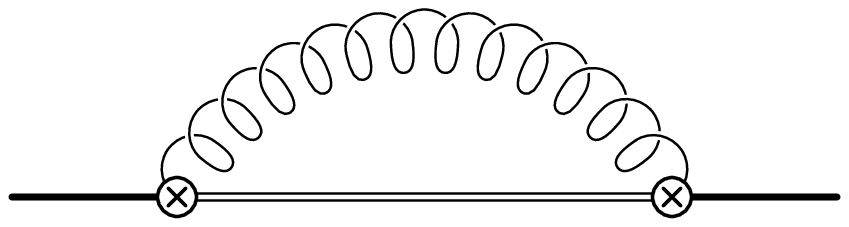}}
\put(73,-3){$\underbrace{\hbox{~~~~~~~~~~~~~~~~~~~~~~~~~}}$}
\put(65,-22){$1/(E-V^{(0)}_o-{\bf p}^2/m)$}
\put(200,1){$\displaystyle{\sim  G_c(E)\int { d^{d}{\bf k} 
\over (2\pi)^{d}}
{\bf r}\,{k \over k+{\bf p}^2/m+V^{(0)}_o-E}
{\bf r}\,G_c(E)}$}
\end{figure}
$$
\sim G_c(E)\,
{\bf r}\, ({\bf p}^2/m+V^{(0)}_o-E)^3
\left\{
{1 \over \epsilon}+\gamma+\ln{({\bf p}^2/m+V^{(0)}_o-E)^2 \over \nu_{us}^2}+C
\right\}\,
{\bf r}\, G_c(E)
\,,
$$
which also produces divergences that get absorbed in $\delta h_s$. Overall, we get a 
consistent EFT.

\end{document}